\documentclass[letterpaper, 10 pt, conference]{ieeeconf}

\IEEEoverridecommandlockouts                              

\usepackage{mathtools}
\usepackage{mathrsfs}

\usepackage{cite}
\usepackage{soul}
\usepackage{lipsum}
\usepackage{amsmath} 
\usepackage{amssymb}  
\usepackage{amsfonts}
\usepackage{mathrsfs}
\usepackage{mathtools}
\usepackage{xcolor}
\usepackage[thmmarks, amsmath]{ntheorem}

\title{\LARGE \bf
Active Front-End Rectifier Modeling and Analysis
}

\author{Aidar Zhetessov and Nathan Petersen
}

\begin{document}

\maketitle
\thispagestyle{plain}
\pagestyle{plain}

\begin{abstract}
Modeling and comprehensive analysis of the Active Front-End (AFE) rectifier is addressed in this work. The particular emphasis is on the comparison of two state feedback control frameworks: state-space and frequency domain. The equivalence of both frameworks for the AFE case was shown. Moreover, analytical relations between control gains and system parameters was derived, allowing for explicit gain scheduling as a function of operating point over the mission profile. The state observation of grid currents has also been addressed. Finally, the work shows the example of robustness evaluation in presence of system variation using the Lyapunov functions approach. The system in nominal conditions was shown to be robust with respect to $50\%$ AC hardware parameter variations.
\end{abstract}

\section{Introduction}

The power electronic supply of high-power electrical systems from the three-phase ac mains is usually carried out using three-phase rectifier systems. Unlike passive rectifiers, Active Front-End (AFE) rectifiers utilize their active components and controls to regulate the output dc voltage to the desired values while keeping high quality of the power drawn from the grid \cite{Kolar}. In addition to establishing a smooth grid-load power electronic interface, the same AFE topology can be used to provide further grid services like reactive power compensation and, more recently, even primary grid-forming control \cite{Zhong}.

Proper operation of the AFE in mentioned applications depends on the control structure employed with AFE. Moreover, as it was shown in \cite{Shi}, not only the operation, but even the AFE hardware design itself might depend on the employed control structure for a given set of performance requirements. Namely, \cite{Shi} has found the regulator-induced constraints on electrical parameters of the rectifier ($L, C, f_{sw}$). These constraints limit the design search space of the rectifier, while guaranteeing the prescribed performance requirements. Overall, AFE rectifier control influences hardware design/optimization as well as proper operation in various classical and emerging applications. Thus, it is interesting to perform a comprehensive analysis of the AFE rectifier controls from different perspectives.

This project addresses the AFE rectifier modeling, control, and state observation aspects in two frameworks: state-space and frequency domain. The state-space approach to the control and observation problems, taught in ECE717, is to be juxtaposed to the frequency domain approach, undertaken in \cite{Shi} and ME746. By doing this, one expects to gain a better understanding of the two control frameworks and some insights that each of these perspectives can provide using the example of the AFE rectifier system.

The rest of the project is organized as follows: Section \ref{Model} derives the AFE rectifier small-signal model of interest. The control and observation problems are analyzed and discussed from different perspectives in Sections \ref{Control} and \ref{Observe}, respectively. The robust evaluation of the system under parameter variations is addressed in Section \ref{Lyapunov}. Finally, the work is summarized in Section \ref{Conclusions}.

\section{AFE Rectifier Modeling} \label{Model}

\begin{figure}
    \centering
    \includegraphics[width=\columnwidth]{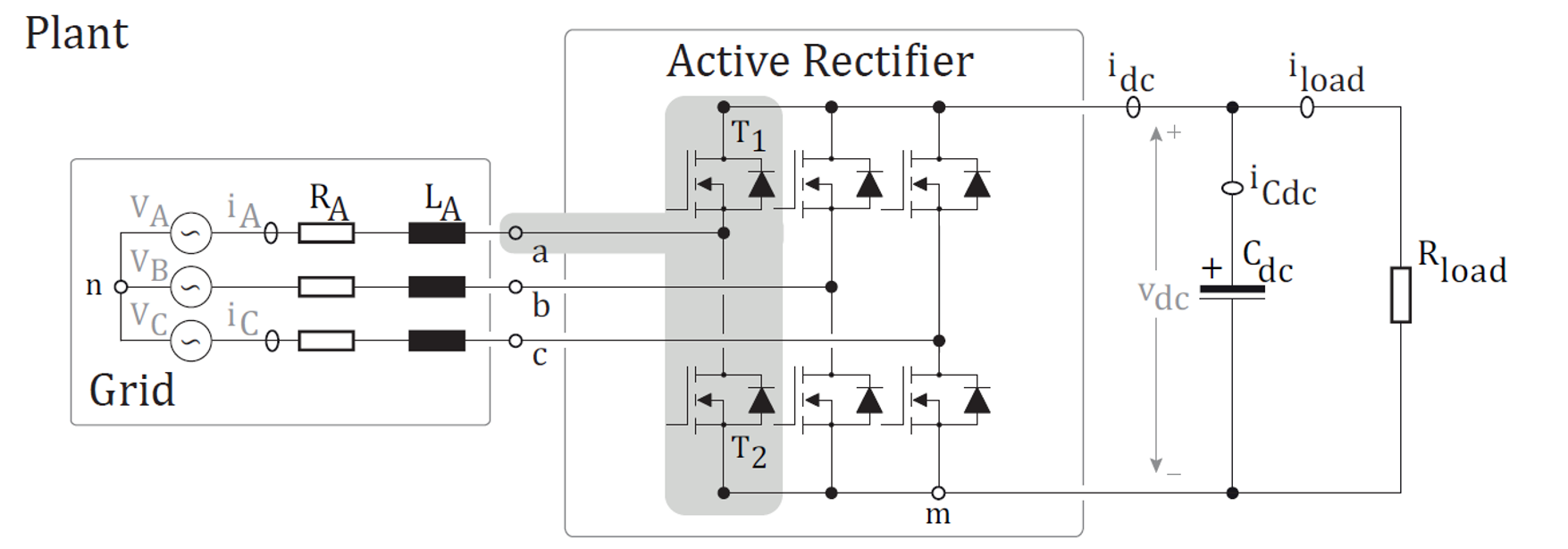}
    \caption{Active Front End Rectifier Schematic Diagram}
    \label{fig:AFE}
\end{figure}

The proposed idea was to model, analyze, and control the grid-connected three-phase AFE rectifier, realized as a standard six-switch two-level three-phase topology. The rectifier can be modeled to feed the resistive, constant current/power, or full ZIP-load. An example schematic of the rectifier with resistive load is shown in Fig. \ref{fig:AFE}.
Here, the power comes from the grid, which is denoted by the balanced three-phase voltage sources $V_A, V_B, V_C$. The three-phase grid current passes through primary passive filter components ($L$ and $R$ same for three phases) to enter the switching stage of the rectifier. The switching stage transfers AC currents to DC side as prescribed by the control law of the rectifier (yet to be defined). The converted DC current $i_{dc}$ feeds the DC-link capacitor $C$ and resistive load $R_{load}$ as per Fig. \ref{fig:AFE}. 

The initial nonlinear model and electrical parameters are taken from \cite{Shi}. Table \ref{t:Specs} summarizes the respective specifications.
\begin{table}[h]
    \caption{System Specifications}
    \begin{center}
        \begin{tabular}{||c |c |c |c ||} 
         \hline
         Parameter & Symbol & Value & Unit \\ 
         \hline\hline
         AC Input LL RMS Voltage & $V_{gLL}$ & 230 & Vrms\\
         \hline
         AC Input Ph Pk Voltage & $V_{g}$ & 187.8 & V\\
         \hline
         DC-bus Voltage & $V_{dc}$ & 400 & V \\
         \hline
         Rated Power & $P$ & 25 & kW\\
         \hline
         Rated Load Resistance & $R$ & 6.4 & $\Omega$\\
         \hline
         Switching Frequency & $f_{sw}$ & 10 & kHz\\ 
         \hline
         Grid Frequency & $f_{o}$ & 60 & Hz\\ 
         \hline
         Input Inductance & $L$ & 0.34 & mH\\ 
         \hline
         DC-bus Capacitance & $C$ & 1300 & uF\\ 
         \hline
         $L$ Series Resistance & $r$ & 5 & m$\Omega$\\ 
         \hline\hline
         Operating Point d-Current & $I_{gd}$ & 88.96 & A\\ 
         \hline
         Operating Point q-Current & $I_{gq}$ & 0 & A\\ 
         \hline
         Operating Point d-Duty & $M_d$ & 0.4684 & -\\
         \hline
         Operating Point q-Duty & $M_q$ & -0.0285 & -\\
         \hline
         Common-Mode Duty & $M_{CM}$ & 0.5 & -\\
         \hline\hline
         Current Control Bandwidth & $f_{i}$ & 1 & kHz\\ 
         \hline
         Voltage Control Bandwidth & $f_{v}$ & 100 & Hz\\ 
         \hline
        \end{tabular}
        \label{t:Specs}
    \end{center}
\end{table}
The nonlinear large-signal model in \cite{Shi} comes from the averaged voltage/current equations written in the dq-frame aligned with the grid voltage ($v_{gd}$):
\begin{equation}
\begin{split}
&L\frac{i_{gd}}{dt} = v_{gd} - m_d v_{dc} - i_{gd} r + \omega_{0} L i_{gq}
\\
&L\frac{i_{gq}}{dt} = v_{gq} - m_q v_{dc} - i_{gq} r - \omega_{0} L i_{gd}
\\
&C\frac{v_{dc}}{dt} = \frac{3}{2} \big(m_d i_{gd} + m_q i_{gq}\big) - \frac{v_{dc}}{R}
\\
\end{split}
\label{eq:nonlinplant}
\end{equation}
After perturbing/linearizing the nonlinear plant equations (\ref{eq:nonlinplant}) one can arrive at the following steady-state operating point relations \eqref{eq:OPrelations} and state-space plant model \eqref{eq:nonlinplant}:
\begin{equation}
\begin{split}
M_d V_{dc} &= V_{gd} - I_{gd}r
\\
M_q V_{dc} &= - \omega_{0} L I_{gd}
\\
P = \frac{V_{dc}^2}{R} &=\frac{3}{2} \big(M_d V_{dc} I_{gd}\big)
\\
\end{split}
\label{eq:OPrelations}
\end{equation}
\begin{equation}
\begin{split}
&\underbrace{\begin{bmatrix} \dot{\Tilde{i}}_{gd} \\ \dot{\Tilde{i}}_{gq} \\ \dot{\Tilde{v}}_{dc} \end{bmatrix}}_{\mathbf{\dot{x}}}
= 
\underbrace{\begin{bmatrix}
    -\frac{r}{L} & \omega_0 & -\frac{M_d}{L} \\
    -\omega_0 & -\frac{r}{L} & -\frac{M_q}{L} \\
    \frac{3M_d}{2C} & \frac{3M_q}{2C} & -\frac{1}{CR}
\end{bmatrix}}_{\mathbf{A}} \underbrace{\begin{bmatrix} \Tilde{i}_{gd} \\ \Tilde{i}_{gq} \\ \Tilde{v}_{dc} \end{bmatrix}}_{\mathbf{x}}
+ \\
&\underbrace{\begin{bmatrix}
    -\frac{V_{dc}}{L} & 0 \\
    0 & -\frac{V_{dc}}{L} \\
    \frac{3I_{gd}}{2C} & \frac{3I_{gq}}{2C}
\end{bmatrix}}_{\mathbf{B_1}} \underbrace{\begin{bmatrix} \Tilde{m}_d \\ \Tilde{m}_q \end{bmatrix}}_{\mathbf{u_1}} 
+
\underbrace{\begin{bmatrix}
    \frac{1}{L} & 0 \\
    0 & \frac{1}{L} \\
    0 & 0
\end{bmatrix}}_{\mathbf{B_2}} \underbrace{\begin{bmatrix} \Tilde{v}_{gd} \\ \Tilde{v}_{gq} \end{bmatrix}}_{\mathbf{u_2}} \\
&\underbrace{\begin{bmatrix} \Tilde{v}_{dc} \end{bmatrix}}_{\mathbf{y}}
=
\underbrace{\begin{bmatrix}
    0 & 0 & 1 \\
\end{bmatrix}}_{\mathbf{C}} \underbrace{\begin{bmatrix} \Tilde{i}_{gd} \\ \Tilde{i}_{gq} \\ \Tilde{v}_{dc} \end{bmatrix}}_{\mathbf{x}}, \mathbf{D} = \mathbf{0}
\end{split}
\label{eq:linplant}
\end{equation}
Here the state variables were selected to be small-signal grid dq-frame currents and DC voltage. The input $B$ matrix was split in two parts to differentiate between manipulated inputs (small-signal modulation indices) and disturbances (small-signal grid voltage disturbances). Output is one of the state variables to be controlled - small-signal DC voltage. The other state variable of interest is the small-signal q-axis grid current. $M_d$, $M_q$, $I_{gd}$, $I_{gq}=0$, $V_{dc}$, $V_{gd}$ and $V_{gq}=0$ are the operating point large-signal parameters, around which the linearization was performed. At rated conditions and in dq-frame aligned with grid $v_{gd}$ voltage, these steady state parameters are evaluated in Table \ref{t:Specs}. Note that the common-mode duty cycle $M_{CM}$ reflects the steady state voltage between the grid star point and the lower rail of the rectifier. For a standard PWM modulation assumed in this work, the mentioned voltage is kept constant at $V_{CM} = M_{CM} V_{dc} = 0.5 V_{dc}$.

\section{AFE Control in Two Frameworks: State-Space and Frequency Domain} \label{Control}

This section investigates the State Feedback (SFB) Control applied to the AFE rectifier and looks at the control problem from two perspectives: State-Space (SS, ECE717) and Frequency Domain (FD, ME746). Originally, full state feedback control method pertains to the linear systems theory and state-space framework taught in linear systems classes like ECE717. On the other hand, there is so-called "full state feedback" approach to the control in the frequency domain framework taught in ME746 controls class, originating from the work of Prof. Lorenz at UW-Madison. The objective of this section is to compare/contrast the two perspectives and conclude whether ``full state feedback" control in frequency domain (FD) is actually implementing the same control law as the original state feedback control in state-space (SS). AFE rectifier serves as the plant to be controlled in both cases, with the modeling addressed in Section \ref{Model}. Note that since this section addresses control problem only, it is assumed that the full state of the system is known and the grid voltage disturbances are known and negligible. Subsequently we relax some of these assumptions in the context of observation problem. The other assumptions are: perfect knowledge of the system matrices, perfect system modeling via these matrices, and the constant linearized operating point.

This section comprises four subsections:

\begin{enumerate}
  \item \textbf{Controllability Analysis}: a prerequisite to the control design of the AFE plant.
  \item \textbf{State-Space SFB Controller Design}: conventional approach to full state feedback control.
  \item \textbf{Frequency Domain SFB Controller Design}: ``physics-based" approach to full state feedback control.
  \item \textbf{Comparison of SS \& FD SFBs}: representation of FD SFB block diagram in equivalent SS SFB matrix form, vice versa, and comparison.
\end{enumerate}

\subsection{Controllability Analysis}

\begin{figure*}[ht]
    \centering
    \makebox[\textwidth][c]{\includegraphics[width=1.20\textwidth]{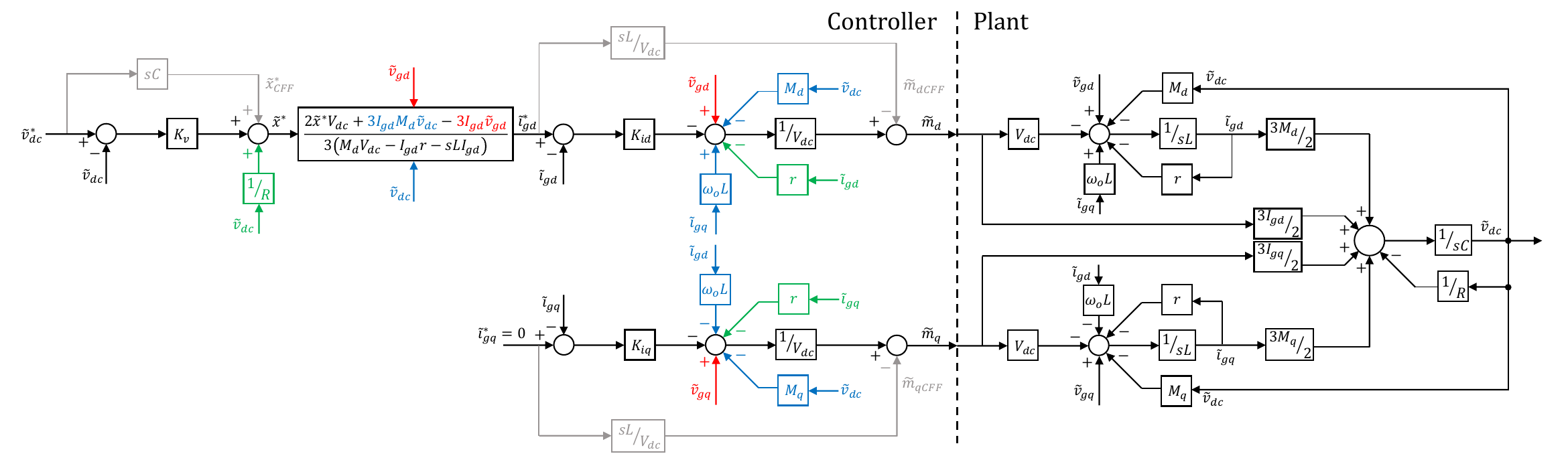}}
    \caption{Frequency domain state feedback controller design using the techniques from ME746: Disturbance Input Decoupling \textcolor{red}{(DID)}, State Feedback Decoupling \textcolor{cyan}{(SFBD)}, Virtual Zero Reference Decoupling \textcolor{green}{(VZRD)}, Command Feedforward \textcolor{lightgray}{(CFF)}.}
    \label{fig:FD_SFBC}
\end{figure*}

From linear system theory, the controllability of the system can be checked by controllability matrix:
\begin{align}
\mathcal{C}_\text{CTR} = \begin{bmatrix}
\mathbf{B} & \mathbf{AB} & \cdots & \mathbf{A}^{n-1}\mathbf{B}
\end{bmatrix}
\end{align}
and if $\text{rank} (\mathcal{C}_\text{CTR}) = n$, the system is called controllable. This means that the state vector $\mathbf{x}$ can be steered back to the origin from any initial state $\mathbf{x}_0 \in \mathbb{R}^n$.

Plugging in $n=3$ and the matrices $\mathbf{A}$ and $\mathbf{B}_1$ from \eqref{eq:linplant}, the controllability matrix is computed as:
\begin{align}
\mathbf{\mathcal{C}}_\text{CTR} = \begin{bmatrix}
-\frac{V_{dc}}{L} & 0 & \frac{3 I_{gd}}{2 C} \\
0 & -\frac{V_{dc}}{L} & \frac{3 I_{gq}}{2 C} \\
\mathcal{C}_{\text{CTR},1,3} & \mathcal{C}_{\text{CTR},2,3} & \mathcal{C}_{\text{CTR},3,3}  \\
\mathcal{C}_{\text{CTR},1,4} & \mathcal{C}_{\text{CTR},2,4} & \mathcal{C}_{\text{CTR},3,4}  \\
\mathcal{C}_{\text{CTR},1,5} & \mathcal{C}_{\text{CTR},2,5} & \mathcal{C}_{\text{CTR},3,5} \\
\mathcal{C}_{\text{CTR},1,6} & \mathcal{C}_{\text{CTR},2,6} & \mathcal{C}_{\text{CTR},3,6} \\ \label{eq:c_ctr}
\end{bmatrix}^\top
\end{align}
where the third column entries are derived as follows:
\begin{equation}
\begin{split}
\mathcal{C}_{\text{CTR},1,3} &= \frac{r V_{dc}}{L^2} - \frac{3 I_{gd} M_d}{2 C L} \\
\mathcal{C}_{\text{CTR},2,3} &= \frac{\omega_0 V_{dc}}{L} - \frac{3 I_{gd} M_q}{2 C L} \\
\mathcal{C}_{\text{CTR},3,3} &= -\frac{3 I_{gd}}{2 C^2 R} - \frac{3 M_d V_{dc}}{2 C L}
\end{split}
\label{eq:c_ctr13} 
\end{equation}
The rest of the entries are not included due to their size. Checking the rank of $\mathcal{C}_\text{CTR}$ from \eqref{eq:c_ctr} using MATLAB, it is indeed 3, meaning that the system is controllable.

\subsection{State-Space State Feedback Controller Design}

Provided controllability of the system, the state feedback controller can place the closed-loop eigenvalues to any desired locations. Generally, the desired eigenvalue locations are selected as close to the real axis and as far to the left-half-plane of the real-imaginary plane as possible. This is to ensure faster convergence, less overshoots, less oscillations during transients, and greater disturbance attenuation/rejection. On the other hand, the eigenvalue locations are limited by the system saturation/slew rate limits, noise attenuation/rejection specifications, and model inaccuracies, which tend to become more pronounced at higher frequencies and set the constraints on system robustness. 

Within the guidelines above, state feedback allows placing eigenvalues near or far from each other. This quite different from the conventional FD control design methods, which often utilize time scale separation of the cascaded control loops. It would be interesting to investigate how does the state feedback eigenvalue placement impact the manipulated input  requirements, transients etc. This work, though, is limited only to the comparison of two frameworks of state feedback control design. Therefore, for both SS SFB and FD SFB controllers the conventional AFE eigenvalue locations are selected.

Given the switching frequency of $10$ kHz from Table \ref{t:Specs}, within the conventional setting, two inner current loop closed-loop poles are chosen to be at $1$ kHz ($\lambda_{1,2} = -2\pi f_i = -2\pi \cdot 1000$ Hz). The outer voltage loop pole is selected to be at $100$ Hz ($\lambda_{3} = -2\pi f_v = -2\pi \cdot 100$ Hz) due to a decade frequency separation between the loops. These poles are to be fixed for both controllers of this work.

Taking into account all the assumptions above, the state dynamics from \eqref{eq:linplant} with SS SFB controller take the following form:

\begin{equation}
\mathbf{\dot{x}} = \mathbf{A}\mathbf{x} + \mathbf{B_1} \underbrace{(-\mathbf{K_{ss}}\mathbf{x})}_{\mathbf{u_1}}
\label{eq:SS_SFB}
\end{equation}

Here $\mathbf{K_{ss}} \in \mathbb{R}^{2 \times 3}$ is a SS SFB controller matrix that moves the closed-loop eigenvalues of the state dynamics to the desired locations. Using the \textsc{Place} command in MATLAB one can find $\mathbf{K_{ss}}$ provided $\mathbf{A}$, $\mathbf{B_1}$, and the list of desired eigenvalues $\mathbf{\Lambda} = [\lambda_1, \lambda_2, \lambda_3]$: 

\begin{equation}
\begin{split}
    \mathbf{K_{ss}} &= \textsc{Place}(\mathbf{A},\mathbf{B_1},[\lambda_1, \lambda_2, \lambda_3]) = \\
    &= \begin{bmatrix}
    -5.9682 & -0.3204 & -2.3864  \\
    0.3204 & -5.3283 & -0.0713  \\
    \end{bmatrix} \cdot 10^{-3}
\end{split}
\label{eq:SS_SFB_K}
\end{equation}

Note that for multi-input systems the algorithm uses extra degrees of freedom to find a solution $\mathbf{K}$ that minimizes the sensitivity of closed-loop poles to perturbations in $\mathbf{A}$ and $\mathbf{B_1}$ \cite{Kautsky}. Obtained SS SFB controller is to be compared/contrasted to its FD SFB counterpart.

\subsection{Frequency Domain State Feedback Controller Design}

The classical AFE frequency domain controller is a cascaded design with inner current and outer voltage loops, as in \cite{Shi}. For such cascaded designs, additional "physics-based" FD techniques were proposed in ME746 course, which result in so-called "state feedback" (SFB) controller, just like the conventional SFB in state-space framework. Although these techniques do not nominally change the cascaded control structure, they end-up eliminating additional control states associated with integrators, making the control structure depend only on the feedback states. The original SS SFB control also does not introduce additional states, hence the similarity and the name is deduced. Generally, there are practical reasons why one might want to keep the cascaded control structure as opposed to explicit SS SFB configuration and vice versa. Those will be touched upon in the next subsection. In this subsection, starting from the control structure in \cite{Shi} and applying the aforementioned techniques from ME746, one derives the FD SFB controller design for the AFE in Fig. \ref{fig:AFE} and \eqref{eq:linplant}.

The classical cascaded controller (in black) with "physics-based" modifications (in color) for the AFE plant is depicted in the closed-loop block diagram of Fig. \ref{fig:FD_SFBC}. To the right of the dashed vertical line one can find the detailed block diagram of the AFE plant \eqref{eq:linplant}. To the left of the dashed vertical line one can find the controller. The integrator gains were removed from the conventional controller in black due to the modifications in color. Classical structure uses d- and q-manipulated inputs to regulate d- and q-current components respectively. The q-component current reference is zero, while the d-component current reference is dictated by the outer cascaded DC voltage controller.

Four frequency domain techniques from ME746 were used to bring the classical controller to the "SFB" form: Disturbance Input Decoupling (\textcolor{red}{DID}), State Feedback Decoupling (\textcolor{cyan}{SFBD}), Virtual Zero Reference Decoupling (\textcolor{green}{VZRD}), and the Command Feedforward (\textcolor{lightgray}{CFF}). DID (red in Fig. \ref{fig:FD_SFBC}) attempts to cancel the external disturbances by measuring and applying the same and opposite signals through manipulated inputs. In the AFE case, grid voltage small-signal disturbances could be decoupled as shown. SFBD (blue in Fig. \ref{fig:FD_SFBC}) attempts to cancel the influence of all other states on the given state dynamics by measuring and applying the same and opposite signals through manipulated inputs. For example, from the plant in can be seen that the $\Tilde{i}_{gd}$ dynamics are influenced by the $\Tilde{v}_{dc}$ through $M_d$ and the relevant SFBD decouples this effect. VZRD (green in Fig. \ref{fig:FD_SFBC}) can be regarded as the subset of SFBD, which accounts for the influence of the state itself on its own dynamics. Together, DID, SFBD, and VZRD break down the plant dynamics such that it becomes the first order system, which can be easily controlled with zero steady state error using only the proportional gain. For example, focusing only on the $\Tilde{i}_{gd}$ dynamics, DID, SFBD, and VZRD cancel all the influences of the plant on $\Tilde{i}_{gd}$, leaving simple $1/sL$ as the plant to be controlled. By choosing the gain $K_{id} = \omega_i L$, one simply places the closed-loop pole of the inner $\Tilde{i}_{gd}$ loop to the desired location $\omega_i$. Finally, CFF (gray in Fig. \ref{fig:FD_SFBC}) accounts for potentially non-constant commands into the respective loops and feeds those forward to the manipulated inputs to ensure the best command tracking (no phase lag and signal attenuation).

Once all the control modifications are made, one can tune the respective cascaded loops to the desired bandwidths that match SS SFB controller and Table \ref{t:Specs}. For the q-axis current loop gain:

\begin{equation}
    T_q(s) = \frac{K_{iq}}{sL} \rightarrow K_{iq} = 2\pi f_i L = \omega_i L
\label{eq:Gq}
\end{equation}

For the d-axis cascaded current and voltage closed-loop transfer function:

\begin{equation}
\begin{split}
    G_{d}(s) &= \frac{1}{\frac{s^2 C L}{K_{id} K_v} + \frac{s C}{K_v} + 1} = \frac{1}{\frac{s^2}{\omega_i \omega_v} + \frac{s(\omega_v+\omega_i)}{\omega_i \omega_v} + 1} \\
    \rightarrow K_{id} &= (\omega_i + \omega_v)L,\quad  K_v = \frac{\omega_i\omega_v C}{(\omega_i + \omega_v)} \\
\end{split}
\label{eq:Gd}
\end{equation}

Finally, by setting the $\Tilde{v}^*_{dc} = 0$ and multiplying out all the gains for dc-conditions (since both references are at zero) one arrives at the following equations for the manipulated inputs:

\begin{equation}
\begin{split}
    \Tilde{m}_d &= -k_{11} \Tilde{i}_{gd} -k_{12} \Tilde{i}_{gq} -k_{13} \Tilde{v}_{dc} \\
    \Tilde{m}_q &= -k_{21} \Tilde{i}_{gd} -k_{22} \Tilde{i}_{gq} -k_{23} \Tilde{v}_{dc} \\
\end{split}
\label{eq:mdq}
\end{equation}

Where the gains $k_{ij}$ are as follows:

\begin{equation}
\begin{split}
    k_{11} &= \frac{r-K_{id}}{V_{dc}} = -5.8623 \cdot 10^{-3} \\
    k_{12} &= -\frac{\omega_o L}{V_{dc}} = -0.3204 \cdot 10^{-3} \\
    k_{13} &= \frac{M_d}{V_{dc}} - \frac{2K_{id}V_{dc}[K_v-R^{-1}]-3K_{id}I_{gd}M_d}{3V_{dc}[M_dV_{dc}-I_{gd}r]} = \\
    &= -2.4338 \cdot 10^{-3} \\
    k_{21} &= \frac{\omega_o L}{V_{dc}} = 0.3204 \cdot 10^{-3} \\
    k_{22} &= \frac{K_{iq}-r}{V_{dc}} = -5.3282 \cdot 10^{-3} \\
    k_{23} &= \frac{M_q}{V_{dc}} = -0.0713 \cdot 10^{-3} \\
\end{split}
\label{eq:gainsK}
\end{equation}

Or, after collecting together in the matrix $\mathbf{K_{fd}}$:

\begin{equation}
    \mathbf{K_{fd}} = \begin{bmatrix}
    -5.8623 & -0.3204 & -2.4338  \\
    0.3204 & -5.3282 & -0.0713  \\
    \end{bmatrix} \cdot 10^{-3}
\label{eq:FD_SFB_K}
\end{equation}

Obtained FD SFB controller in matrix form is to be compared to its SS SFB counterpart.

\subsection{Comparison of State-Space \& Frequency Domain SFBs}

Comparing the controllers in \eqref{eq:SS_SFB_K} and \eqref{eq:FD_SFB_K} one can see that they are within 2\% difference from each other in the worst case for the operating point at hand. Other operating points were also evaluated, and the observed alignment holds true for those points as well. Therefore, at least for the AFE case, one can conclude that "physics-based" full state feedback control in frequency domain does implement the original state feedback control in state space and that the "state feedback" term usage in this case is correct.

Moreover, one can claim that the designed FD SFB controller has the highest robustness to parameter variations in $\mathbf{A}$ and $\mathbf{B_1}$, since it is the practically the same as the SS SFB controller, for which the robustness claim is true due to \textsc{Place} command and \cite{Kautsky}. More on robustness evaluation can be found in Section \ref{Lyapunov}.

Comparing the design effort of the controllers, state space approach is more attractive. Indeed, the state space approach facilitates fast and precise closed-loop pole placement, proven robustness guarantees \cite{Kautsky}, and easy command feedforward realization without the need to consider cascaded loop interactions. 

Comparing the practical realization aspects of the controllers, frequency domain is more appealing. Indeed, even without additional controller states, it uses cascading, which  is intuitively useful to separate the low-energy fast state dynamics from high-energy slow state dynamics. Although in the case of AFE this separation is not that obvious, at least from operating point evaluation the capacitive energy storage $\frac{CV_{dc}^2}{2} = 104J$ is far greater than inductive energy storage $\frac{LI_{gd}^2}{2} = 1.34J$. From the safety perspective, it is useful to have an explicit current loop, on which the current reference limiter can be applied. From the discrete control perspective, faster state (current) can be sampled at much higher frequency than slower state (voltage), implying less computational hardware allocation requirements for the outer loops computations. Also, lower sampling frequency for the outer states increases the resolution per sample. From the mission profile operation perspective, the fact that FD SFB controller provides explicit analytic relations \eqref{eq:gainsK} between controller gains and the operating points is very useful, as one can implement gain scheduling with variable gains for secure operation at different operating points over the mission profile. On the other hand, the SS SFB controller $\mathbf{K_{ss}}$ relation to the parameters in $\mathbf{A}$ and $\mathbf{B_1}$ is not that obvious. Finally, time scale separating the inner and outer loops makes system more robust. Qualitatively, when the eigenvalues are far from each other, it takes a lot of parameter variation to significantly change the closed-loop dynamics, while close eigenvalues can move/break off real axis with some parameter variation, significantly changing the nature of the dynamics. More on robustness is in Section \ref{Lyapunov}.

\section{AFE State Observer} \label{Observe}

The system formulation \eqref{eq:linplant} is such that only voltage sensors are used for feedback. This means that the dc voltage $\Tilde{v}_{dc}$ is measured, but also the dq-frame grid voltages $\Tilde{v}_{gd}$ and $\Tilde{v}_{gq}$ are measured. In \eqref{eq:linplant}, the grid voltages are contained in the input $\mathbf{u}_2$; they are not a state variable. To implement control of the AFE, the state vector is required including the currents $\Tilde{i}_{gd}$ and $\Tilde{i}_{gq}$. Since the currents are not measured directly, they must be estimated.

In this section, the observability of the AFE is investigated. Then, an observer structure is proposed to estimate the unmeasured currents.

\subsection{Observability Analysis}
\label{sec:obs_analysis}

From linear system theory, controllability is related to observability through the dual system formulation. In the end, the observability matrix is defined as
\begin{align}
\mathcal{O}_\text{OBS} = \begin{bmatrix}
\mathbf{C} \\
\mathbf{CA} \\
\vdots \\
\mathbf{CA}^{n-1}
\end{bmatrix}
\end{align}
and if $\text{rank} (\mathcal{O}_\text{OBS}) = n$, the system is called observable. This means that the state vector $\mathbf{x}$ can be inferred from knowledge of the output $y$ and input $u$, and the system matrices.

Plugging in $n=3$ and the matrices $\mathbf{A}$ and $\mathbf{C}$ from \eqref{eq:linplant}, the observability matrix is computed as:
\begin{align}
\mathbf{\mathcal{O}}_\text{OBS} = \begin{bmatrix}
0 & 0 & 1 \\
\frac{3 M_d}{2 C} & \frac{3 M_q}{2 C} & -\frac{1}{R C} \\
\mathcal{O}_{\text{OBS},3,1} & \mathcal{O}_{\text{OBS},3,2} & \mathcal{O}_{\text{OBS},3,3} \\ \label{eq:o_obs}
\end{bmatrix}
\end{align}
where
\begin{align}
\mathcal{O}_{\text{OBS},3,1} &= \frac{-3 R (L M_q \omega_0 + M_d r) C - 3 M_d L}{2 C^2 L R} \nonumber \\
\mathcal{O}_{\text{OBS},3,2} &= \frac{3 R (L M_d \omega_0 - M_q r)C - 3 M_q L}{2 C^2 L R} \nonumber \\
\mathcal{O}_{\text{OBS},3,3} &= \frac{-3 R^2 (M_d^2 + M_q^2)C + 2 L}{2 C^2 L R^2} \nonumber
\end{align}

The rank of $\mathcal{O}_\text{OBS}$ from \eqref{eq:o_obs} is indeed 3, meaning that the system is observable.

\subsection{Observer Design}

\begin{figure}
    \centering
    \includegraphics[width=0.9\columnwidth]{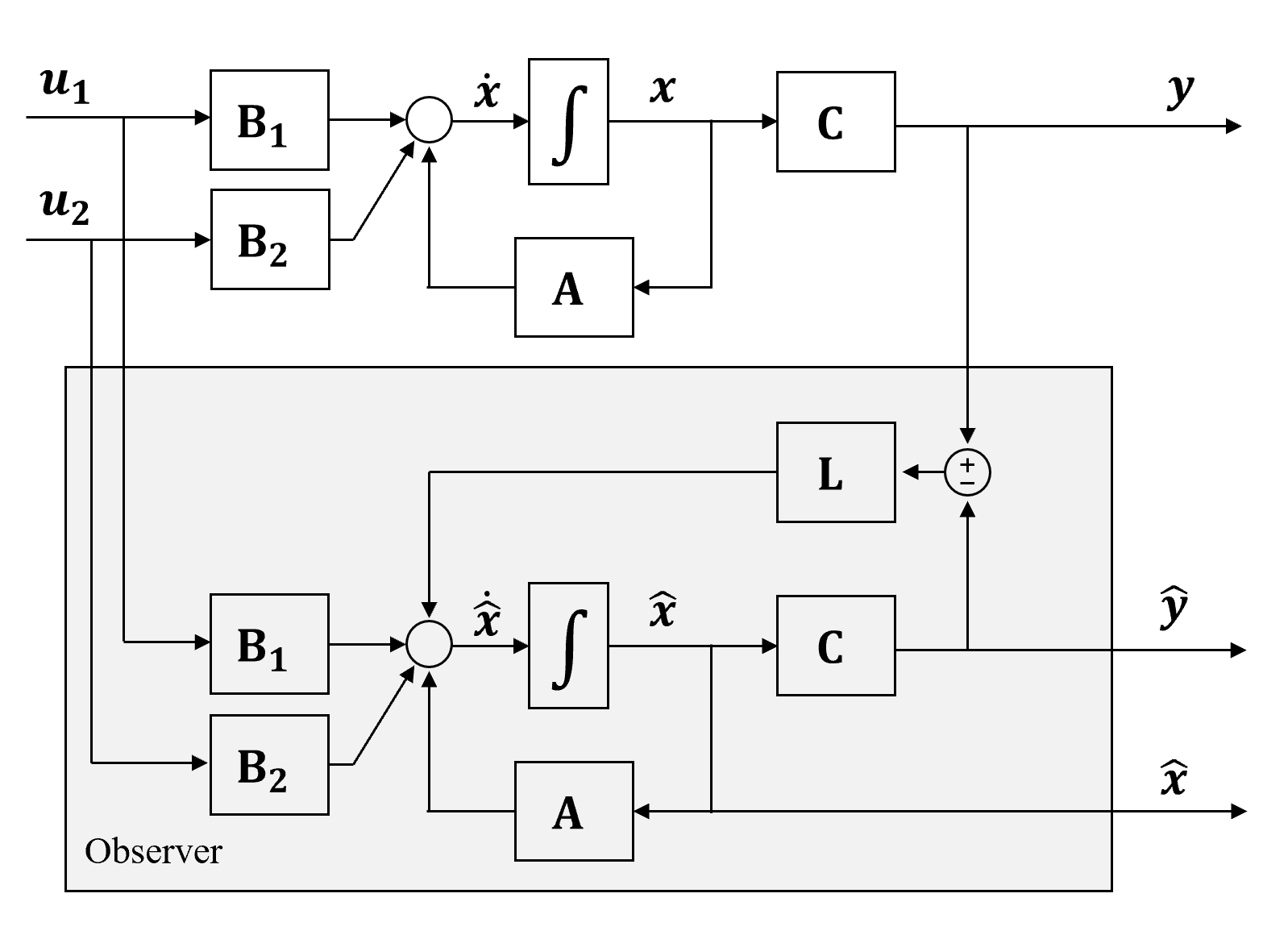}
    \caption{Proposed observer structure to estimate the unmeasured currents, $\Tilde{i}_{gd}$ and $\Tilde{i}_{gq}$, which are contained in $\hat{\mathbf{x}}$.}
    \label{fig:observer_diagram}
\end{figure}

Since the system is observable per Section~\ref{sec:obs_analysis}, an observer can be constructed which can accurately estimate the unmeasured states: $\Tilde{i}_{gd}$ and $\Tilde{i}_{gq}$. The observer is formed using the system matrices from \eqref{eq:linplant}: $\mathbf{A}$, $\mathbf{B}_1$, $\mathbf{B}_2$, and $\mathbf{C}$. It is assumed that all matrices are known perfectly and that the real system ideally follows the linear model. Furthermore, the operating point at which the system is linearized must be constant.

The block diagram of the proposed observer is shown in Fig.~\ref{fig:observer_diagram} using the notation from \eqref{eq:linplant}. The input to the observer is split into two parts: (i) the control input: $\mathbf{u}_1$, and (ii) the measured voltages: $\mathbf{u}_2$ and $\mathbf{y}$. The feedback matrix gain $\mathbf{L}$ is used to tune the performance of the observer, and thus the estimation convergence properties.

\subsection{Observer Tuning}

The observer feedback gain $\mathbf{L}$ is critical to the estimation performance. The observer dynamics are governed by:
\begin{align}
    \dot{\hat{\mathbf{x}}} &= \mathbf{A} \hat{\mathbf{x}} + \mathbf{B}_1 \mathbf{u}_1 + \mathbf{B}_2 \mathbf{u}_2 + \mathbf{L} (\mathbf{y} - \hat{\mathbf{y}}) \nonumber \\
    \hat{\mathbf{y}} &= \mathbf{C} \hat{\mathbf{x}}
\end{align}
Defining the state estimation error as $\mathbf{e} = \mathbf{x} - \hat{\mathbf{x}}$, the observer dynamics can be written in terms of the $\mathbf{e}$ as:
\begin{align}
    \dot{\mathbf{e}} = (\mathbf{A} - \mathbf{L}\mathbf{C}) \mathbf{e} \label{eq:obs_err_dynamics}
\end{align}
The dynamics of the estimation error \eqref{eq:obs_err_dynamics} can be arbitrarily set by placing the eigenvalues of $\mathbf{A} - \mathbf{L}\mathbf{C}$ via selecting $\mathbf{L}$. Recall that this is possible because the system is \textit{observable}. Typically, the state estimation dynamics are selected to be at least an order of magnitude faster than the control dynamics.

\section{Robustness Evaluation through Lyapunov Functions} \label{Lyapunov}

Throughout the work there were several mentions of system robustness with respect to parameter variations. MATLAB \textsc{Place} command uses the eigenvalue sensitivity to $\mathbf{A}$ and $\mathbf{B_1}$ parameter variations as the robustness representation. Qualitative discussion on FD SFB controller mentions closed-loop pole separation as the sign of a robust system. To evaluate system robustness more precisely, one can use the Lyapunov stability approach to the time-varying systems.

For a closed-loop time varying systems with certain bounds on parameter variations in $\mathbf{A_{cl}}(t)$, one can conclude asymptotic stability of the system irrespective of the temporal nature of parameter variations within the defined bounds, if there exists a positive-definite Lyapunov function $V(\mathbf{x})$ with a negative-definite time derivative for all possible $\mathbf{A_{cl}}(t)$.

Consider $50\%$ parameter variations in $L$ and $r$ due to thermal/saturation/proximity effects for example. Since $L,r$ enter the system matrices $\mathbf{A}$ and $\mathbf{B_1}$, entries of those will change, affecting $\mathbf{A_{cl}}(t)=\mathbf{A}(t)-\mathbf{B_1}(t)\mathbf{K}$. As those variations are not modeled, the controller matrix will not change (both SS and FD SFB). Robust asymptotic stability of the system in such conditions can be concluded if there exists $\mathbf{P}=\mathbf{P}^\top>0$ such that:

\begin{equation}
\begin{split}
    V(\mathbf{x}) &= \mathbf{x}^\top \mathbf{P} \mathbf{x} > 0 \\
    \dot{V}(\mathbf{x}) &= \mathbf{x}^\top \big(\mathbf{A_{cl}}^\top(t) \mathbf{P} + \mathbf{P} \mathbf{A_{cl}}(t)\big) \mathbf{x} < 0, \forall \mathbf{x}\neq 0, \forall t \\
\end{split}
\label{eq:Lyapunov}
\end{equation}

To solve for $\mathbf{P}$, one can define $\mathbf{Q} = \mathbf{I}$, equate $\dot{V}(\mathbf{x}) = -\mathbf{Q}$, and use the MATLAB \textsc{Lyap} command to find $\mathbf{P}$. The resulting matrix form served as the inspiration to come up with the following energy-like matrix $\mathbf{P}$:

\begin{equation}
    \mathbf{P} = \begin{bmatrix}
\frac{L}{2} & 0 & \frac{\sqrt{LC}}{4} \\
0 & \frac{L}{2} & 0 \\
\frac{\sqrt{LC}}{4} & 0 & \frac{C}{2} \\
\end{bmatrix} = \mathbf{P}^\top> 0
\label{eq:P}
\end{equation}

\begin{figure}
    \centering
    \includegraphics[width=0.9\columnwidth]{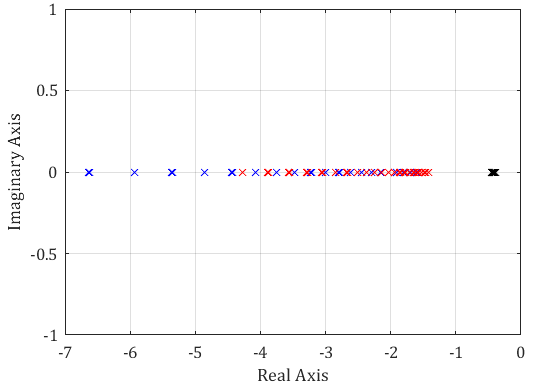}
    \caption{Eigenvalue sweep of the $\mathbf{A_{cl}}^\top(t) \mathbf{P} + \mathbf{P} \mathbf{A_{cl}}(t)$ matrix.}
    \label{fig:Evalues}
\end{figure}

The range of parameters of interest ($0.5L \rightarrow 1.5L, 0.5r \rightarrow 1.5r$) was swept with high resolution to evaluate the $\mathbf{A_{cl}}(t)$ in each case, assuming $\mathbf{K}$ is constant. For each case, negative-definiteness of $\mathbf{A_{cl}}^\top(t) \mathbf{P} + \mathbf{P} \mathbf{A_{cl}}(t)$ was evaluated by checking the respective eigenvalues. Here, if all eigenvalues throughout the sweep are strictly negative, one can conclude robust asymptotic stability of the system in presence of parameter variation. The respective eigenvalues are depicted in Fig. \ref{fig:Evalues}.

Thus, it was shown that even with AC-side parameter variations in $\pm50\%$ range, the closed-loop system is still robustly asymptotically stable. Note that such conclusions cannot be drawn simply from evaluating the closed-loop eigenvalues of the system over the parameter range. Lyapunov functions offer a powerful tools for analysis and control design. Mastering those tools further is definitely an interesting outlook for the future.

\section{Conclusions} \label{Conclusions}

This work addressed modeling and comprehensive analysis of the Active Front-End (AFE) rectifier system from two perspectives: state-space and frequency domain. First, the tight hardware-control interconnection was mentioned in the introduction, motivating a deeper understanding of the AFE rectifier control for hardware design and overall system operation. Next, small-signal linearized model has been derived in Section \ref{Model}. 

Using the derived model, Section \ref{Control} compared two state feedback control frameworks - state-space and frequency domain. It turned out that, at least for the AFE case, the "physics-based" frequency domain SFB controller is actually the original state-space SFB controller, so the term "state feedback" for FD SFB controller was applied correctly. In the process, one could identify the analytical relations between SFB controller gains and the system parameters, including operating points, which can be useful for mission profile regulation.

Next, section \ref{Observe} addressed the AFE state observer design, considering only voltage measurements being available. This can be quite useful from the practical perspective, as the current sensors require not only sensors themselves, but all the accompanying peripherals like traces, a/d conversion, layout space etc.

Finally, the Section \ref{Lyapunov} showed how the Lyapunov functions can be used for robustness evaluation of the time varying system under certain parameter constrained variation conditions. It turns out that there exists a Lyapunov function that can prove system robust asymptotic stability in the presence of $50\%$ variation in AC hardware parameters.

\section*{Contributions}

The AFE Modeling, Control in Two Frameworks, and Robustness Evaluation were addressed by Aidar Zhetessov. The AFE State Observation was addressed by Nathan Petersen. Other sections were addressed in collaboration.

\bibliographystyle{IEEEtran}
\bibliography{mybibfile}

\end{document}